\documentclass[sigconf,nonacm]{acmart}  
\setcopyright{none}
\AtBeginDocument{%
  }

\usepackage{makecell}
\usepackage{booktabs}

\usepackage{graphicx}
\usepackage{array} 
\usepackage{multirow}
\usepackage{booktabs}  
\usepackage{float}

\begin{document}

\title{Can LLMs Generate Behaviors for Embodied Virtual Agents Based on Personality Traits?}

\author{Bin Han}

\affiliation{%
  \institution{University of Southern California}
  \city{Los Angeles}
  \state{CA}
  \country{USA}
}
\email{binhan@usc.edu}

\author{Deuksin Kwon}

\affiliation{%
  \institution{University of Southern California}
  \city{Los Angeles}
  \state{CA}
  \country{USA}
}
\email{deuksink@usc.edu}

\author{Spencer Lin}

\affiliation{%
  \institution{University of Southern California}
  \city{Los Angeles}
  \state{CA}
  \country{USA}
}
\email{linspenc@usc.edu}

\author{Kaleen Shrestha}

\affiliation{%
  \institution{University of Southern California}
  \city{Los Angeles}
  \state{CA}
  \country{USA}
}
\email{kshresth@usc.edu}

\author{Jonathan Gratch}

\affiliation{%
  \institution{University of Southern California}
  \city{Los Angeles}
  \state{CA}
  \country{USA}
}
\email{gratch@ict.usc.edu}

%
\renewcommand{\shortauthors}{Han et al.}

\begin{abstract}
This study proposes a framework that uses personality prompting with Large Language Models (LLMs) to generate verbal and non-verbal behaviors for virtual agents based on personality traits. 
Focusing on extraversion, we evaluated the system across two scenarios—negotiation and ice-breaking—using both introverted and extroverted agents.
In Experiment 1, we ran agent–agent simulations and conducted linguistic analysis and personality classification to assess whether the LLM-generated language reflected the intended traits, and whether the corresponding nonverbal behaviors differed by personality.
In Experiment 2, we conducted a user study to evaluate whether these personality-aligned behaviors were consistent with their intended traits and perceptible to human observers.
Our results show that LLMs can generate verbal and nonverbal behaviors that align with personality traits, and that users are able to recognize these traits through the agents’ behaviors.
This work highlights the potential of LLMs in shaping personality-aligned virtual agents.
\end{abstract}



\maketitle

\section{Introduction}

Large Language Models (LLMs) have increasingly been used across various domains, and recent advancements in multimodal LLMs have enabled the processing and generation of not only text but also images and videos~\cite{cui2024survey, caffagni2024revolution}. These models go beyond conventional natural language processing, integrating multi-modal information to enhance understanding and interaction capabilities~\cite{wang2024comprehensive}.

\begin{figure}
    \centering
\includegraphics[width=1\linewidth]{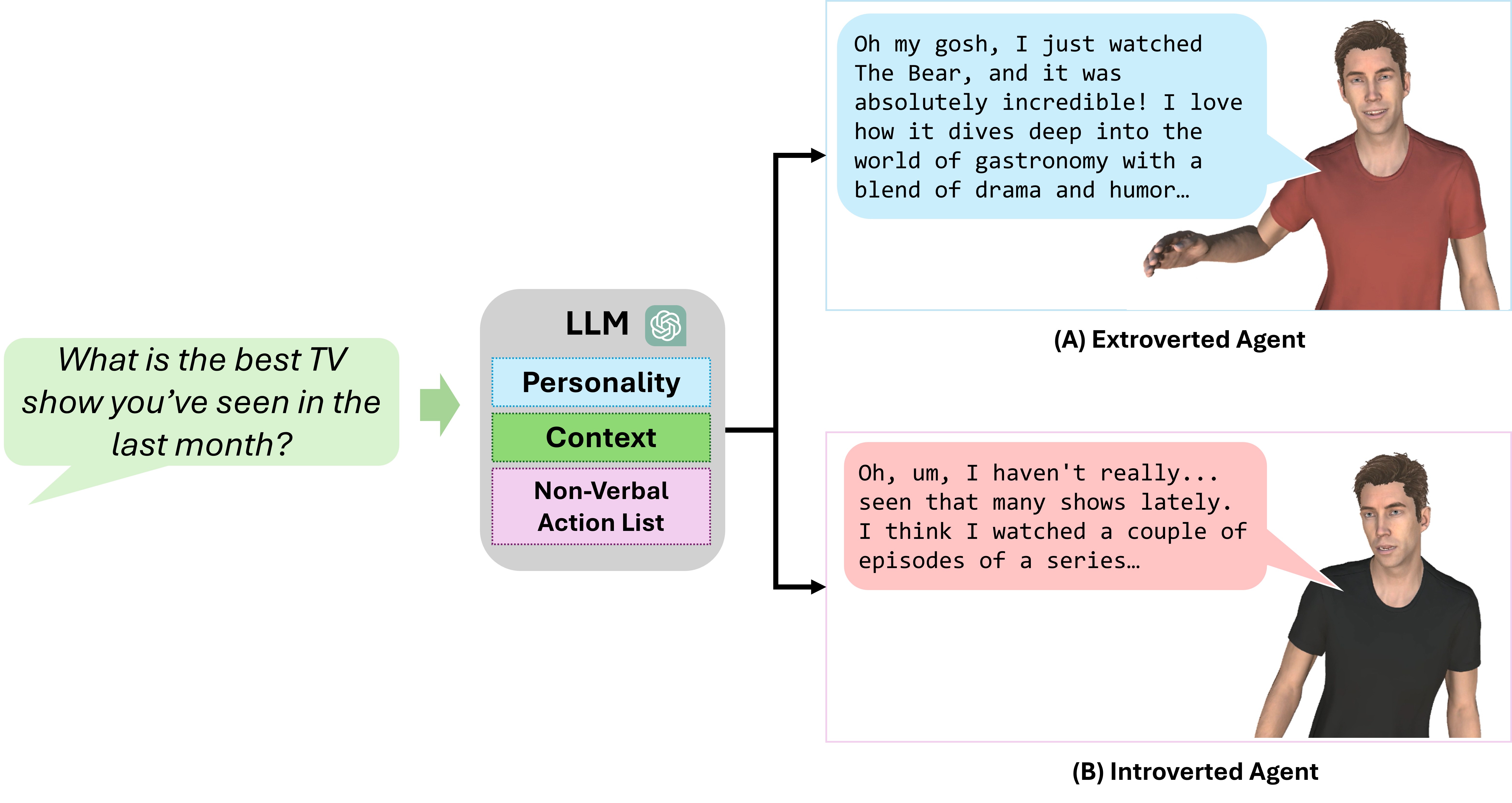}
    \caption{Conversational system overview behind the introverted and extroverted virtual agent verbal and nonverbal behavior generation.}
    \label{fig:enter-label}
\end{figure}

Virtual Agents are increasingly incorporating LLMs to enhance interaction capabilities (e.g., through adaptive dialogue, emotional understanding)~\cite{lin2023toward, hale2024integration}. 
Previously, many Virtual Agents relied on rule-based approaches, where responses were generated using predefined scripts~\cite{reithinger2006virtualhuman,devault2014simsensei,kim2024engaged}.
To move beyond rigid scripting, some early work explored machine learning techniques to generate personality-aligned dialogue or linguistic styles~\cite{mairesse2011controlling}, but such approaches often required task-specific datasets.
More recently, LLMs have emerged as a flexible alternative, enabling adaptive dialogue management that can handle diverse conversational contexts without the need for extensive domain-specific training~\cite{soliman2024using, arjmand2024empathic}. 
As a result, LLMs are now being widely adopted across various domains for improving Virtual Agent interactions~\cite{steenstra2024virtual, wan2024building}.
Furthermore, LLMs have been showing promising potential in simulating personality traits and generating personality-aligned responses. 
Some studies focus on evaluating the extent to which LLMs inherently reflect stable personality characteristics~\cite{serapio2023personality}, while others propose prompting strategies to induce target traits during generation~\cite{jiang2023evaluating}.

Building on this progress, we consider whether LLMs can serve not only as expressive agents, but as controllable virtual confederates in social interactions—interactive systems that consistently reflect specific personality traits across tasks. 
These agents can be carefully manipulated to examine how behavioral traits influence social outcomes, such as whether users collaborate more effectively with partners who match their personality~\cite{blascovich2011infinite,de2011effect}
Earlier systems achieving this level of control required extensive hand-crafting and were limited to specific tasks. In contrast, LLMs now show remarkable flexibility across conversational tasks (e.g., negotiation~\cite{hale2024integration}, tutoring), and growing evidence suggests they can be guided to adopt psychological states (e.g., anger~\cite{arjmand2024empathic}) or traits (e.g., extroversion~\cite{serapio2023personality}). 
However, such demonstrations have largely been restricted to text. 
This work explores whether LLMs can generate coherent verbal and nonverbal behaviors that consistently reflect personality traits.

LLMs have been widely used in Virtual Agent systems, particularly for generating and managing verbal behaviors such as dialogue. Recently, however, there has been a growing interest in leveraging LLMs for nonverbal behavior generation as well~\cite{torshizi2025large}, aiming to create more contextually adaptive agents. Earlier work, in contrast, primarily focused on nonverbal behavior generation using rule-based approaches~\cite{devault2014simsensei} or style-matching methods based on real-time facial mimicry~\cite{hoegen2019end}, which, while controllable, often lacked flexibility across diverse conversational situations.

With the rapid evolution of multimodal LLMs, it is now possible to explore whether these models can generate nonverbal behaviors in a more dynamic and context-aware manner. 
Early studies that leverage LLMs for empathetic nonverbal cue generation in social robots~\cite{lee2023developing} suggest that these models have shown the potential to generate nonverbal expressions alongside verbal responses.

Our study focuses on leveraging LLMs to generate both verbal and nonverbal behaviors that align with personality traits, specifically extraversion.
Extraversion is a suitable trait for this exploration because it is strongly associated with observable and measurable behaviors, including facial expressions, gestures, and vocal characteristics~\cite{neff2010evaluating, lippa1998nonverbal, buck1974sex}.
Moreover, aligning a virtual agent’s behavior with specific personality traits—particularly along the extraversion dimension—has been shown to enhance user satisfaction, perceived friendliness, and engagement in both social and task-oriented interactions~\cite{jin2023birds, tapus2008user, isbister2000consistency}.
By including both extroverted and introverted agents, we aim to assess whether LLMs can generate personality-consistent behaviors and how these behaviors influence user perception across different contexts.

To investigate this, we address the following research questions:
\begin{itemize}
    \item \textbf{RQ1:}  Can LLMs generate behaviors for virtual agents that align with established patterns of extraversion in the literature?
    \item \textbf{RQ2:} Do humans accurately perceive personality traits in Virtual Agents based on their verbal and nonverbal behaviors generated by LLMs?
    \item \textbf{RQ3:} How do verbal and nonverbal behaviors contribute to the perception of personality in Virtual Agents, and which modality (verbal vs. nonverbal) has a stronger influence?
\end{itemize}

By addressing these questions, this research aims to provide insights into the role of LLMs in generating personality-aligned behaviors for Virtual Agents and their impact on user perception.

\section{Related Work}

\subsection{Personality and Behavior}


Personality refers to individual differences in patterns of thoughts, emotions, and behaviors that are consistently exhibited over time. 
One of the most widely accepted frameworks for describing personality is the Big Five model, also known as OCEAN, which includes Openness to Experience, Conscientiousness, Extraversion, Agreeableness, and Neuroticism~\cite{goldberg2013alternative}.

Individuals high in Openness to Experience are more imaginative, curious, and open to novel experiences. McCrae~\cite{mccrae1987creativity} found that Openness to Experience is closely associated with creative problem-solving and the use of symbolic language.
Those high in Conscientiousness tend to be organized, self-disciplined, and goal-oriented. According to Barrick and Mount~\cite{barrick1991big}, conscientious individuals often perform better in workplace settings due to their structured and reliable behaviors.
Agreeable individuals are characterized by their cooperative and empathetic nature. 
Graziano and Eisenberg~\cite{graziano1997agreeableness} showed that high agreeableness reduces interpersonal conflict and facilitates positive social interactions.

Personality is also clearly reflected in nonverbal behaviors. Rutter et al.~\cite{rutter1972visual} found that extroverts are more likely to maintain eye contact during group discussions, whereas introverts tend to avert their gaze. Buck et al.~\cite{buck1974sex} showed that extroverts display more dynamic and expressive facial expressions, while introverts exhibit more subtle and restrained behavior.
However, how personality is expressed through behavior can vary depending on the situation. Some contexts make it easier for people to show their personality, while others may constrain it~\cite{funder2006towards}. 
This is often described using the “traffic light” analogy. Some situations resemble green lights and encourage expression. Others resemble red lights and inhibit it. To explore this idea, we compare negotiation and ice-breaking contexts, as they differ in how freely personality is expressed.

Several studies suggest that an observer’s own personality traits can influence how they perceive others' personalities. In particular, some individuals are more consistent and sensitive in interpreting personality cues, a phenomenon often referred to as being a “good judge” of personality~\cite{letzring2008good}. Traits such as openness and extraversion have been linked to greater accuracy and sensitivity in personality perception. 
\vspace{-1em}

\subsection{Personality with LLMs}

Based on the ability of LLMs to effectively mimic various human behaviors, cognitive traits, and psychological traits, there is a growing body of research exploring and measuring intrinsic personality traits of LLMs, as well as studies aimed at inducing specific target personality traits in LLMs \cite{sorokovikova2024llms, jiang2023evaluating, serapiogarcía2023personality,lee2024llms}. 

In particular, inducing personality for modern LLMs is a nontrivial maneuver given the temperamental nature of prompting and the goal to induce a consistent personality across user studies. 
Serapio-García et al.~\cite{serapiogarcía2023personality} presented a method based on prompting that synthesizes and effectively evaluates single or multiple personality traits in LLMs, such as the PaLM family. 
Jiang et al.~\cite{jiang2023evaluating} devised an automatic method, named P2, that uses sequential prompting strategies to induce diverse LLM personalities, integrating psychological traits with LLM knowledge. The approaches proposed in both studies are controllable, consistent, and verifiable. 

Recently, there has been growing attention on prompting personality expressions in LLMs. For example, PersonalityEdit~\cite{mao2024editing} proposes a novel task that modifies LLM responses to opinion-based questions in order to reflect specific personality traits. 
Sorokovikova et al.~\cite{sorokovikova2024llms} provide further evidence that LLMs naturally simulate Big Five traits. By prompting models, models like GPT-4, LLaMA-2, and Claude consistently exhibit specific personality profiles without explicit conditioning.  

\subsection{Virtual Agents and Nonverbal Behavior}

Various studies have explored how to control and express personality in virtual agents. 
Isbister and Nass~\cite{isbister2000consistency} created introverted and extroverted characters by manipulating both verbal and nonverbal behaviors. Extroverted characters used strong and friendly language, confident phrasing, and expressive nonverbal cues, whereas introverted characters relied on weaker sentence structures, question-based dialogue, and limited nonverbal expression.
Subsequently, Neff et al.~\cite{neff2010evaluating} systematically analyzed how nonverbal cues such as posture, gestures, and head movements influence users’ perception of an agent’s personality. Their findings showed that extroverted agents tend to use broader and faster gestures, lean forward, and display more frequent movements.

Building on this foundation, several follow-up studies have examined how a virtual agent’s personality influences user perception~\cite{wang2023extrovert, saberi2021expressing}.
Some approaches utilize physical behavior analysis techniques such as Laban Movement Analysis (LMA) to categorize gesture styles associated with different personality traits~\cite{durupinar2016perform, sonlu2021conversational}. 
More recently, Sonlu et al.~\cite{sonlu2025effects} explored how personality expression and embodiment affect user perception in LLM-based conversational agents by integrating a dialogue system powered by an GPT 3.5 Turbo. They designed agent behaviors based on LMA to reflect high and low personality traits, and found that agents expressing high personality traits were perceived as more engaging.

Many existing virtual agent systems generate verbal and nonverbal behaviors separately~\cite{kim2024engaged,sonlu2021conversational}, which may reduce their flexibility in adapting to dynamic conversational contexts.
To address these limitations, our work proposes a novel approach that leverages LLMs to generate both verbal and nonverbal behaviors in alignment with personality traits.

\subsection{LLM-Based Nonverbal Behavior Generation}

Recent approaches have begun to leverage LLMs for generating nonverbal behaviors that align with verbal content and contextual meaning. 
Torshizi~\cite{torshizi2025large} uses GPT-4 to select co-speech gestures that are semantically relevant and contextually appropriate, improving human-agent interaction quality. 
This gesture selection is achieved through in-context learning, where GPT-4 is provided with utterance–gesture pairs to infer appropriate gestures for novel utterances. 
In a more integrated approach, Lee et al.~\cite{lee2023developing} propose a system for social robots that simultaneously generates verbal and nonverbal empathetic behaviors using prompt-based LLM outputs.
While prior work has addressed verbal and nonverbal behavior generation or personality alignment separately, our approach unifies them in a single LLM-based framework to produce personality-aligned behaviors for embodied conversational virtual agents.

\section{Method}

Our method aims to generate personality-aligned verbal and nonverbal behaviors using Large Language Models (LLMs). 
Prior literature has consistently shown that extroverts tend to exhibit more expressive facial expressions, expansive gestures, and louder, faster speech patterns.
Based on prior work, we compiled a set of observable behaviors—our \textit{Nonverbal Action List}—used in prompts and animation descriptions to guide the LLM’s generation of appropriate nonverbal actions (in table~\ref{tab:nonverbal-list}).
In the following subsections, we describe how we categorized these actions 
, and how each set of actions was derived from prior research on personality-related behaviors.

\subsection{Nonverbal Behavior Markup}
We categorize nonverbal behaviors into three modalities: face, body, and speech based on previous approachs~\cite{matsumoto2016apa,lee2023developing}.

\begin{figure*}[h]
    \centering
\includegraphics[width=0.70\linewidth]{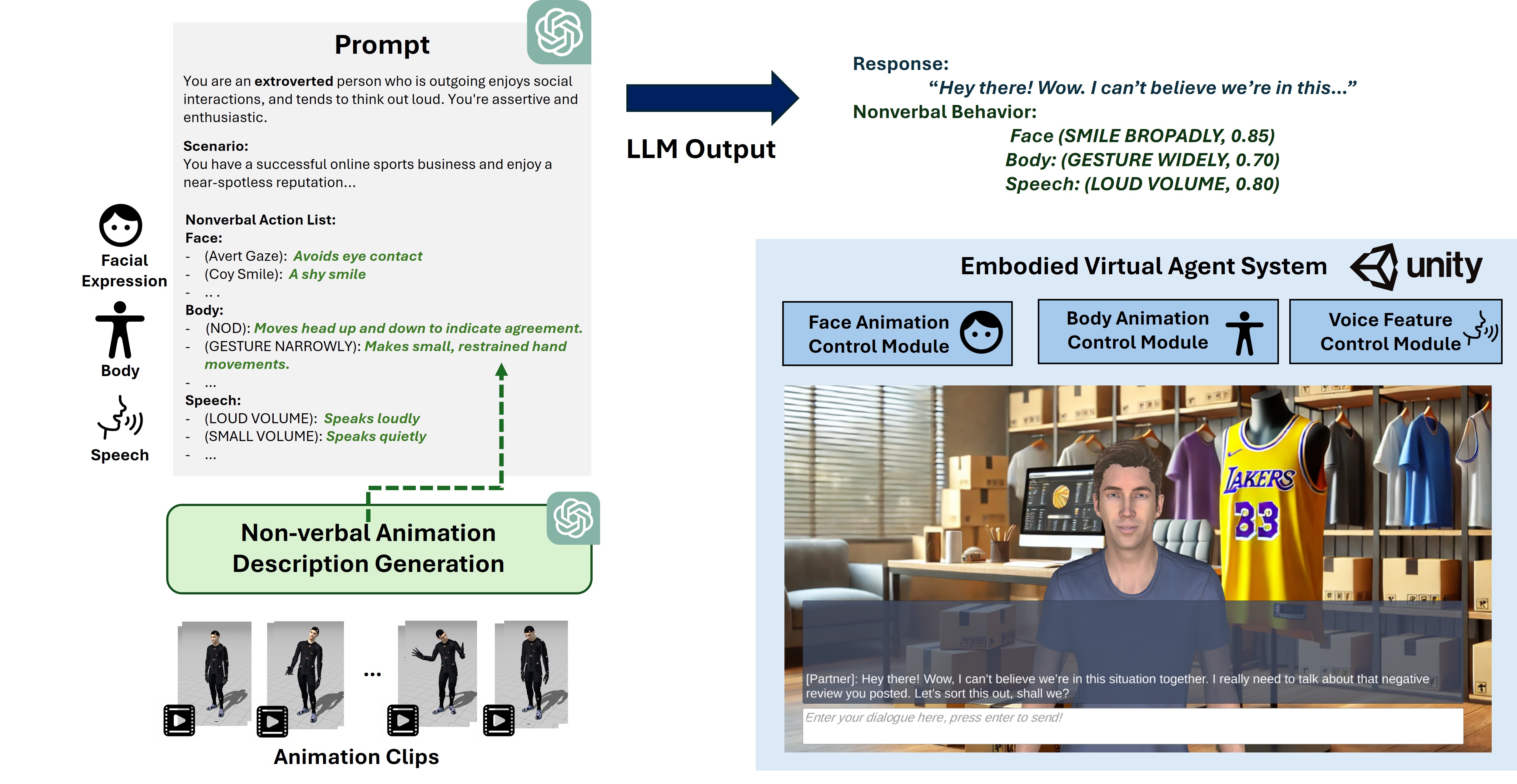}
    \caption{Embodied Virtual Agent System Overview}
    \label{fig:system-overview}
\end{figure*}

\begin{table}[h]
    \centering
    \resizebox{0.45\textwidth}{!}{
    \begin{tabular}{c|c|c}
        \hline
        \hline
        \multicolumn{3}{c}{\textbf{Nonverbal Action List}} \\
        \hline
        \hline
        \textbf{Face} & \textbf{Body} & \textbf{Voice} \\
        \hline
        (Avert Gaze) & (Nod) & (Loud Volume) \\
        (Make Eye Contact) & (Shaking) & (Small Volume) \\
        (Coy Smile) & (Disagree) & (Fast Pace) \\
        (Smile Broadly) & (Agree) & (Slow Pace) \\
        (Subtle Sadness) & (Gesture Narrowly) & \\
        (Intense Sadness) & (Gesture Widely) & \\
        (Mild Anger) & (Gesture Slowly) & \\
        (Strong Anger) & (Gesture Fastly) & \\
        (Soft Surprise) & (Give Thumbs Up) & \\
        (Extreme Surprise) & (Head Tilting) & \\
        (Static Eyebrows) & (Lean Forward) & \\
        (Raise Eyebrows) & (Lean Backward) & \\
        (Narrow Eyes) & & \\
        \hline
        \hline
    \end{tabular}
    }
    \caption{Nonverbal Action List}
    \label{tab:nonverbal-list}
\end{table}

\subsubsection{Facial Expressions}

Facial expressions are crucial in conveying extraversion. Our action list for facial expressions includes 13 distinct labels (see Table~\ref{tab:nonverbal-list}). According to Rutter et al.~\cite{rutter1972visual} and Buck et al.~\cite{buck1974sex}, extroverts tend to display rich and varied facial expressions, while introverts typically exhibit more subdued affect. To reflect this distinction, we include two levels of expressiveness in our facial expression set.
\textit{Subtle expressions}—such as \textit{Coy Smile}, \textit{Subtle Sadness}, \textit{Mild Anger}, and \textit{Soft Surprise}—are designed to reflect the restrained emotional display often associated with introverts. In contrast, \textit{Extreme expressions}—including \textit{Smile Broadly}, \textit{Intense Sadness}, \textit{Strong Anger}, and \textit{Extreme Surprise}—capture the dynamic and exaggerated expressions typical of extroverts.
Furthermore, extroverts are more likely to maintain eye contact during conversations, while introverts tend to avert their gaze~\cite{campbell1978bodily,rutter1972visual}. To represent this behavior, we include the labels \textit{Make Eye Contact} and \textit{Avert Gaze} in the action list.

\subsubsection{Body Movements}

Several studies have shown that extroverts engage in more expansive and frequent gestures~\cite{argyle2013bodily,brebner1985personality}, indicative of an open and dynamic interaction style. 
In contrast, introverts tend to use more restrained and cautious gestures. 
To reflect this distinction, our system includes \textit{Gesture Widely} and \textit{Gesture Fast} to represent the energetic and open behaviors associated with extroversion, while \textit{Gesture Narrowly} and \textit{Gesture Slowly} are used to model the more controlled and reserved behaviors typical of introversion.
Posture also plays a role in signaling personality traits. Extroverts are more likely to adopt a forward-leaning posture, signaling engagement and enthusiasm, whereas introverts tend to lean backward, which can indicate reservation or withdrawal~\cite{lippa1998nonverbal}. Accordingly, we include both \textit{Lean Forward} and \textit{Lean Backward} in our action list to account for these postural cues.
Additional body actions such as \textit{Nod}, \textit{Shaking}, \textit{Agree}, and \textit{Disagree} serve as social feedback cues that can reflect assertiveness and responsiveness—qualities more often associated with extroverts. 

\subsubsection{Voice}

The quality of the voice, such as volume and pace, significantly affect personality perception. Studies by Lee et al.~\cite{lee2021speech} and Scherer~\cite{scherer1978personality} have shown that extroverts tend to speak louder and faster, whereas introverts speak more softly and at a slower pace, reflecting their respective personality traits in communication.
To reflect these differences, we include four voice-related labels in our action list: \textit{Loud Volume}, \textit{Small Volume}, \textit{Fast Pace}, and \textit{Slow Pace}.

\subsection{LLM-driven Behavior Realization}

We implement a behavior realization pipeline that maps LLM-generated output to verbal and nonverbal behaviors in an embodied virtual agent (Figure~\ref{fig:system-overview}).
The system generates both verbal and nonverbal behaviors using LLM and transmits the results to a Unity-based environment via a Flask server.

First, the LLM generates verbal responses based on the given context and personality definition. 
LLM selects appropriate nonverbal behaviors from a predefined action list, ensuring that gestures, facial expressions, and speech characteristics align with the intended personality traits. 
Prompts are defined with the agent's personality and conversational context (e.g., Negotiation, ice-breaking). 
A predefined list of nonverbal actions is included, and the LLM decides descriptions that align with these actions. 
We provide animation descriptions, allowing it to generate contextually appropriate behaviors (e.g., different types of nodding). 
To enable the LLM to understand the available motion repertoire, we develop a \textit{Nonverbal Description Generation Module}. This module uses GPT to extract detailed descriptions of each animation clip in an offline process. For each predefined nonverbal action (e.g., \textit{Wide Gesture}), multiple natural language descriptions are provided (e.g., ``the right hand moves in a large circular motion followed by a simultaneous raise of both arms''). These descriptions serve as interpretable representations of physical movements, enabling the LLM to reason about and select behaviors that align with the intended personality traits. This approach improves the contextual appropriateness and expressive diversity of generated behaviors.

Finally, the selected behaviors are rendered through the embodied virtual agent system developed in Unity, where facial expressions, body movements, and voice features are synchronized to create a coherent and realistic interaction. 
Realistic 3D male character from Reallusion is used (Figure~\ref{fig:system-overview}).
Facial expressions and lip-sync are controlled through the SALSA API, while gesture and posture animations are sourced from Mixamo and Reallusion. 
Speech is generated in real time using the ElevenLabs Text-to-Speech API
To ensure that the selected voices matched the intended personality traits, we conducted a pilot study with 15 participants using four American English voice IDs provided by ElevenLabs.  
Based on participant feedback, the extrovert agent was assigned voice ID \texttt{Eric}, while the introvert agent used voice ID \texttt{Brian}.

Unlike systems such as Lee's NVBG~\cite{lee2006nonverbal} and Pelachaud's Gretta~\cite{bevacqua2010greta}, which synchronize gestures with specific words, our current approach generates nonverbal behaviors at the utterance level. While this allows for coherent overall expression, it lacks the temporal granularity of fine-grained gesture alignment. 
Future work could build upon recent LLM-based methods for word-level gesture generation~\cite{torshizi2025large} to enhance temporal precision.

\section{Experiment 1: Validation for LLM Behavior Generation}
\label{sec:exp1}
To validate the alignment between the personality traits defined in RQ1 and the verbal and nonverbal responses generated by LLMs, we employed an Agent-to-Agent simulation.
In this experiment, personality-specific agents (Introverted and Extroverted) interacted with a Generic agent, which lacks predefined personality traits.

\subsection{Agent-Agent Simulation Setting}

To ensure the generalizability of our findings, we designed two scenarios to validate the system.

\textbf{Negotiation} 
The first scenario involves a buyer-seller conflict adapted from prior work~\cite{hale2025kodis}. 
In this scenario, a misunderstanding drives the interaction, with the buyer seeking a refund and the seller aiming to remove a negative review.
Following prior findings that the buyer allows for more expressive responses, the Personality agent was assigned the buyer, and the Generic agent the seller.

\textbf{Ice-Breaking.} 
The second scenario is designed to create a non-confrontational, emotionally neutral interaction,
adapted from Aron et al.'s Personal Questions~\cite{aron1997experimental}.
The Generic agent asks the Personality agent three ice-breaking questions selected from prior work~\cite{zhang2023ice}
We use GPT\footnote{gpt-4o-mini-2024-07-18, with the temperature set to the default value} as the LLM. 
GPT models are widely adopted in recent research on virtual agents and have demonstrated strong performance in generating coherent and contextually appropriate language and behavior~\cite{torshizi2025large, sonlu2025effects}.
In the Negotiation scenario, we limit conversations to a maximum of 10 turns, while in the Ice-Breaking scenario, we set a limit of 8 turns, as it consists of only three questions.
We conduct 10 trials for each scenario.

\subsection{Behavior Analysis}

We use two text analysis methods to examine the lexical differences between the introverted and extroverted agents. First, we use the widely adopted psycholinguistic text analysis toolkit known as Linguistic Inquiry and Word Count (LIWC)~\cite{boyd2022development}\footnote{LIWC 2015 version}. 
LIWC categorizes words into psychologically meaningful categories. 
LIWC is frequently used to classify the personality of human authors~\cite{koutsoumpis2022kernel}. 
Second, we employ a pre-trained personality detection model~\cite{kazameini2020personality} for Big-5 personality, which uses contextualized embeddings from the BERT language model along with psycholinguistic features for personality trait prediction. The model is used to assess the extent to which the generated responses exhibit measurable differences in extraversion between our agents.
The model outputs a binary result (0 or 1), indicating whether the text is identified as an extrovert.
For nonverbal behavior analysis, we evaluate the predicted probabilities of selected nonverbal actions from a predefined list to identify patterns specific to introverted and extroverted agents. 
We examine if patterns in the likelihood and intensity of nonverbal behaviors reflect the intended personality traits.

\subsection{Result: Experiment 1}

\subsubsection{Verbal Result}
Extroverted agents produced significantly longer and more elaborate responses than introverted agents in both tasks. In the Ice breaking, extrovert verbal utterances had higher word counts (M = 90.58 vs. 56.30) and sentence counts (M = 5.80 vs. 4.92), both $p < .001$. A similar pattern was observed in the Negotiation task (M = 56.32 vs. 41.46, $p < .001$). These findings are consistent with prior literature showing that extroverted individuals tend to produce longer and more elaborate speech~\cite{fast2008personality}

\begin{table}[t]
    \centering
    \caption{LIWC Result - Negotiation Task}
    \label{tab:liwc_negotiation}
    {\scriptsize 
   \begin{tabular}{p{2.6cm} p{0.6cm} p{0.6cm} p{1.2cm} >{\centering\arraybackslash}p{0.5cm} >{\centering\arraybackslash}p{0.8cm}}

        \toprule
        \textbf{LIWC Feature} & \textbf{EXT} & \textbf{INT} & \textbf{t-test} & \textbf{Aligned} & \textbf{Cite} \\
        \midrule
        you (You) & 3.82 & 1.06 & EXT > INT *** & Y & \cite{mairesse2007using} \\
        social (Social) & 5.88 & 2.66 & EXT > INT *** & Y & \cite{tausczik2010psychological} \\
        function (Function Words) & 30.92 & 22.62 & EXT > INT *** & Y & \cite{pennebaker1999linguistic} \\
        pronoun (Pronouns) & 12.74 & 8.62 & EXT > INT *** & Y & \cite{mairesse2007using} \\
        power (Power) & 1.62 & 0.32 & EXT > INT *** & Y & \cite{fast2008personality} \\
        ppron (Personal Pronouns) & 8.42 & 6.12 & EXT > INT *** & Y & \cite{mairesse2007using} \\
        informal (Informal Language) & 0.62 & 0.08 & EXT > INT *** & Y & \cite{pennebaker1999linguistic} \\
        conj (Conjunctions) & 3.20 & 2.12 & EXT > INT *** & Y & \cite{yarkoni2010personality} \\
        certain (Certainty) & 1.06 & 0.50 & EXT > INT *** & Y & \cite{yarkoni2010personality} \\
        negemo (Negative Emotions) & 1.72 & 1.16 & EXT > INT ** & Y & \cite{pennebaker1999linguistic} \\
        nonflu (Nonfluencies) & 0.28 & 0.06 & EXT > INT * & N & \cite{mairesse2007using} \\
        \midrule
        insight (Insight) & 1.34 & 2.86 & EXT < INT *** & Y & \cite{pennebaker1999linguistic} \\
        cogproc (Cognitive Processes) & 6.26 & 8.40 & EXT < INT *** & Y & \cite{pennebaker1999linguistic} \\
        feel (Feel) & 0.08 & 0.46 & EXT < INT *** & Y & \cite{fast2008personality} \\
        tentat (Tentative) & 0.84 & 1.26 & EXT < INT * & Y & \cite{mairesse2007using} \\
        \bottomrule
    \end{tabular}
    }
\end{table}

\begin{table}[t]
    \centering
    \caption{LIWC Result - Ice-breaking Task}
    \label{tab:liwc_ice}
    {\scriptsize
   \begin{tabular}{p{2.6cm} p{0.6cm} p{0.6cm} p{1.2cm} >{\centering\arraybackslash}p{0.5cm} >{\centering\arraybackslash}p{0.8cm}}
        \toprule
        \textbf{LIWC Feature} & \textbf{EXT} & \textbf{INT} & \textbf{t-test} & \textbf{Aligned} & \textbf{Cite} \\
        \midrule
        conj (Conjunctions) & 6.85 & 2.65 & EXT > INT *** & Y & \cite{yarkoni2010personality} \\
        certain (Certainty) & 2.20 & 0.43 & EXT > INT *** & Y & \cite{mairesse2007using} \\
        pronoun (Pronouns) & 16.38 & 11.03 & EXT > INT *** & Y & \cite{mairesse2007using} \\
        social (Social) & 9.53 & 3.90 & EXT > INT *** & Y & \cite{yarkoni2010personality} \\
        affect (Affect) & 9.40 & 4.58 & EXT > INT *** & Y & \cite{pennebaker1999linguistic} \\
        posemo (Positive Emotions) & 8.68 & 4.20 & EXT > INT *** & Y & \cite{pennebaker1999linguistic} \\
        power (Power) & 1.30 & 0.55 & EXT > INT *** & Y & \cite{fast2008personality} \\
        verb (Verbs) & 10.43 & 8.60 & EXT > INT ** & Y & \cite{pennebaker1999linguistic} \\
        \midrule
        nonflu (Nonfluencies) & 1.03 & 2.75 & EXT < INT *** & Y & \cite{mairesse2007using} \\
        informal (Informal Language) & 1.28 & 2.93 & EXT < INT *** & N & \cite{mairesse2007using} \\
        tentat (Tentative) & 1.53 & 2.88 & EXT < INT *** & Y & \cite{mairesse2007using} \\
        \bottomrule
    \end{tabular}
    }
\end{table}

\begin{figure}[t]
    \centering
\includegraphics[width=0.70\linewidth]{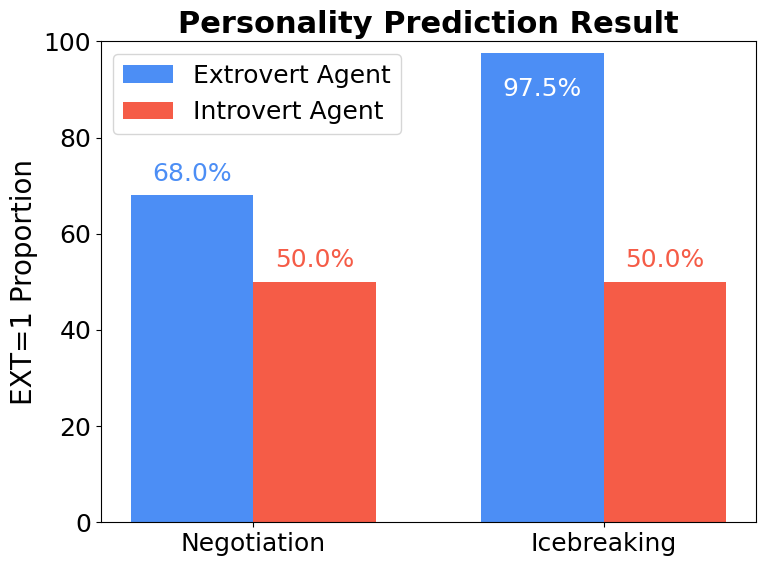}
    \caption{Personality Detection Results~\cite{kazameini2020personality}}
    \label{fig:personality=prediction-model}
\end{figure}

\begin{figure*}[t]
\includegraphics[width=0.8\textwidth]{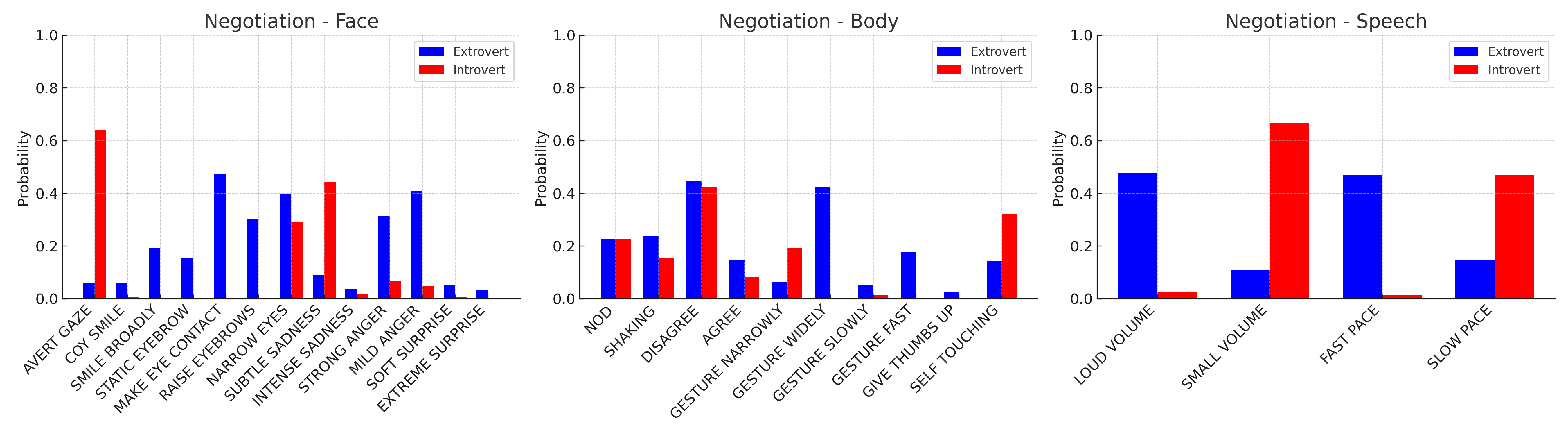}
        \caption{Negotiation- Distribution of nonverbal behavior probabilities across modalities.}
        \label{fig:NV-nego}
\end{figure*}

\begin{figure*}[t]
\includegraphics[width=0.8\textwidth]{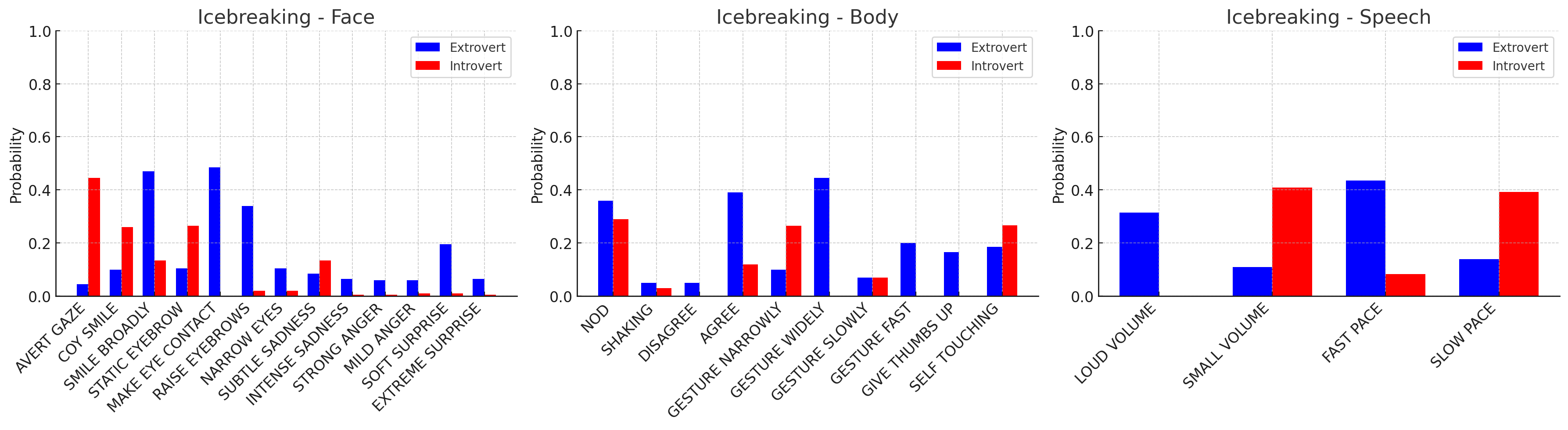}
        \caption{Ice-breaking- Distribution of nonverbal behavior probabilities across modalities.}
        \label{fig:NV-ice}
\end{figure*}

\textbf{LIWC Analysis:} 
The LIWC analysis revealed significant lexical differences between extroverted and introverted agents in both scenarios (see Table~\ref{tab:liwc_ice}, ~\ref{tab:liwc_negotiation}). To ensure both statistical and practical relevance, we report only features that exhibited a significant $t$-test result ($p$ < .05) and a substantial effect size (Cohen’s $d$ > 0.5). This filtering allows us to focus on differences that are both reliable and meaningful. The “Aligned” column indicates whether each observed pattern is consistent with prior literature on linguistic markers of extraversion~\cite{mairesse2007using, yarkoni2010personality, pennebaker1999linguistic, tausczik2010psychological, fast2008personality}. 
Wherever possible, we reference specific studies that report similar findings to support each alignment.
In the Negotiation scenario, 15 LIWC features showed significant differences between extroverted and introverted agents. 
All features except \textit{nonfluencies} were consistent with prior findings on personality-related language use. 
Extroverted agents used more personal pronouns, function words, and social/emotional expressions, whereas introverted agents demonstrated higher usage of insight and cognitive process terms. 
In the Ice-Breaking scenario, 11 LIWC features showed significant differences between extroverted and introverted agents. Among them, all but one \textit{informal (Informal Language)} were consistent with prior literature on personality-related language use. 
Extroverted agents used more social, affective, and power-related words, while introverted agents exhibited greater use of tentative and informal expressions.

\textbf{Perceived Personality Detection:} The personality prediction results show distinct patterns across the two scenarios (see Figure~\ref{fig:personality=prediction-model}).
In the Negotiation scenario, both Extrovert and Introvert Agents had the same proportion (68\%) of utterances classified as extroverted (EXT=1). 
In the Icebreaking scenario, however, the difference became more pronounced. 
The Extrovert Agent's utterances were classified as extroverted 97.5\% of the time, while the Introvert Agent’s utterances were classified as extroverted 50\% of the time. 
This suggests that the Icebreaking context allows for more expressive personality traits, making the distinction between extroverted and introverted behaviors more apparent.

We performed a chi-square test on the influence of scenario and agent type on perceived extroversion. This showed an interaction ($\chi^2(3) = 27.85, p < .001$), which we then examined separately for each scenario. 
Across both scenarios, the classifier was more likely to classify the extroverted agent’s utterances as extroverted, but in the Negotiation scenario, this was only a trend ($\chi^2(1) = 2.65$, $p = .10$), whereas in the Icebreaking scenario, the difference was highly significant ($\chi^2(1) = 20.92$, $p < .001$). 
Together, this suggests personality prompting is successful, but the strength of the manipulation is moderated by the scenario.

\subsubsection{Nonverbal}

As part of our investigation into how LLMs generate personality-aligned nonverbal behaviors, Figure~\ref{fig:NV-nego} and ~\ref{fig:NV-ice} show the average probabilities of each nonverbal action performed by extroverted (blue) and introverted (red) agents across three modalities in two scenarios: Negotiation and Ice-Breaking.

In the Negotiation scenario, extroverted agents displayed a higher probability of expressive facial behaviors such as \textit{Smile Broadly} and \textit{Raise Eyebrows}, while introverted agents showed more restrained cues like \textit{Avert Gaze} and \textit{Subtle Sadness}.
Overall, extroverted agents performed gestures more frequently—particularly \textit{Gesture Widely}—whereas introverted agents were more likely to use \textit{Gesture Narrowly}.
In the voice modality, extroverts tended to use \textit{Fast Pace} and \textit{Loud Volume}, while introverts favored \textit{Slow Pace} and \textit{Small Volume}, consistent with prior findings on vocal expressiveness.
In the Ice-Breaking scenario, the behavioral differences between extroverted and introverted agents were even more pronounced and observable.
Extroverted agents again exhibited more dynamic facial expressions and expansive gestures. The voice patterns followed a consistent trend, with extroverts predominantly using \textit{Fast Pace} and \textit{Loud Volume}, while introverts preferred \textit{Slow Pace} and \textit{Small Volume}.
In addition to personality differences, we also observed task-specific behavioral patterns.
The Negotiation scenario, which involves conflict and disagreement between agents, elicited more anger-related facial expressions and a higher frequency of \textit{Disagree} gestures.
In contrast, the Ice-Breaking scenario, designed to foster positive social interaction, prompted more happiness-related facial expressions and a greater use of \textit{Agree} gestures, reflecting the cooperative and friendly tone of the setting.

\section{Experiment 2: Evaluating Perceived Personality}


Experiment 2 addresses RQ2 and RQ3 by examining whether users can accurately perceive personality traits in embodied virtual agents based on LLM-generated verbal and nonverbal behaviors (RQ2), and by evaluating the relative contribution of each cue to personality perception (RQ3).

\subsection{Experiment Setting}

We conduct a user study using conversations generated through agent-to-agent simulations (in Section~\ref{sec:exp1}). 
Based on these, we create video clips portraying virtual agents with distinct personality traits in two scenarios: Negotiation and Ice-Breaking. 
Each scenario includes interactions where agents exhibit behaviors associated with either extroversion or introversion. Participants are recruited through the online platform Prolific, with 30 individuals assigned to one of the two scenarios (female: 17, male: 13), resulting in a between-subjects design for scenario. 
Within each assigned scenario, participants view two videos: one featuring an extroverted agent and the other an introverted agent. This within-subject manipulation of agent personality enables direct comparison while controlling for scenario-specific context. Additionally, including two different videos—one per agent type—helps reduce potential bias arising from idiosyncratic characteristics of individual videos. 
The study takes approximately 15 minutes, and participants are compensated according to Prolific’s policy.

Participants evaluated the perceived personality of the virtual agents and indicated whether verbal, nonverbal, or both types of behaviors played a greater role in their judgment.
They also selected which specific nonverbal cues—such as facial expressions, posture, gestures, tone of voice, and volume of voice. Multiple selections were allowed.
For verbal behavior, participants chose up to three aspects they found most influential, including speech content, speaking manner, and utterance length.
In addition, participants rated whether the observed behaviors aligned with the agent’s perceived personality traits and were given the opportunity to provide open-ended feedback on mismatches or unnatural expressions.
Participants’ extroversion scores were collected using items 2 and 6 of the BFI-10~\cite{rammstedt2007measuring}. 

\subsection{Result: Experiment 2}

\subsubsection{Participant Personality} 

The mean self-reported extroversion score was 3.8 (SD = 0.96) on a 7-point scale, indicating that participants were generally moderate in extroversion.

\begin{figure} 
    \centering
\includegraphics[width=0.8\linewidth]{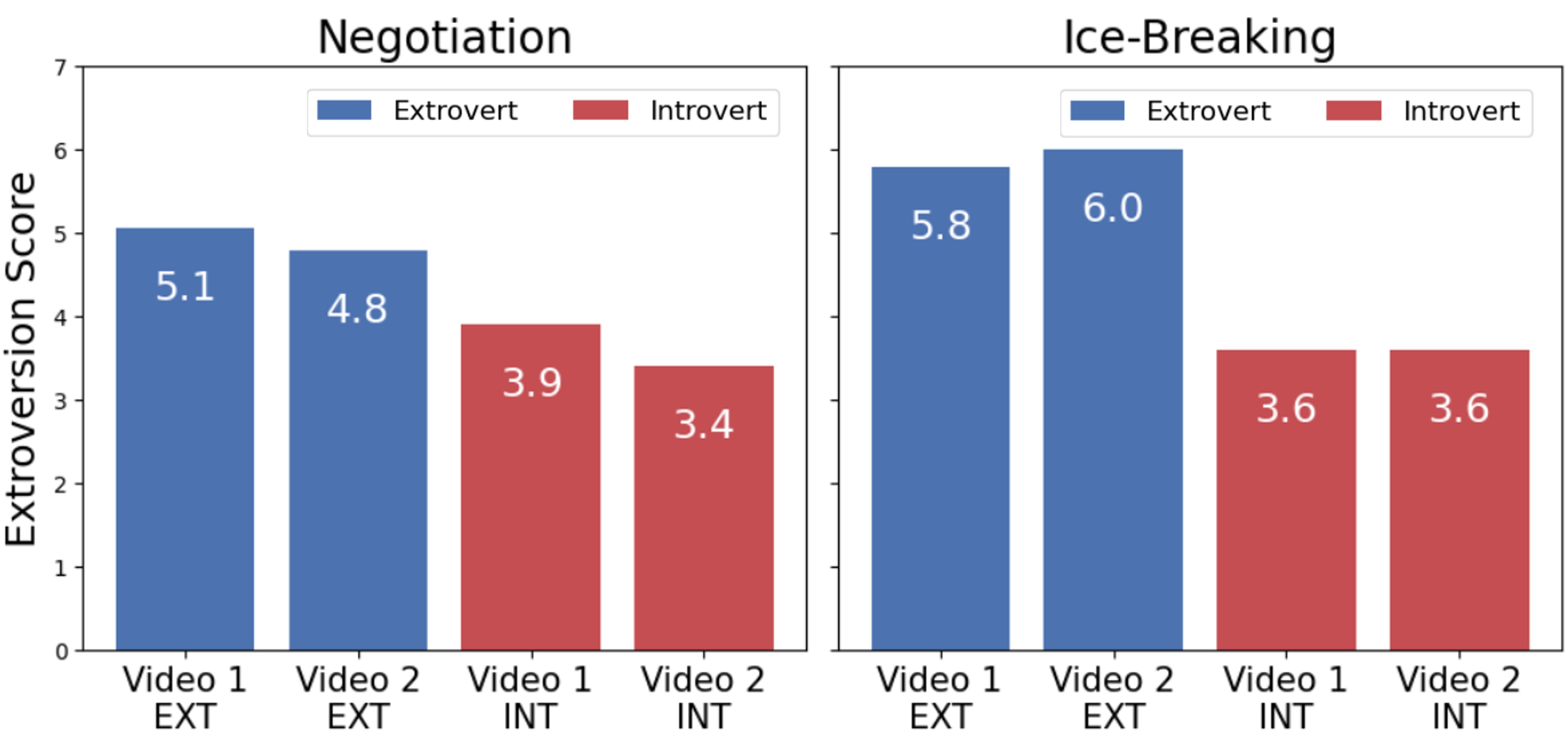}
    \caption{Perceived Extroversion Score}
    \label{fig:user-study-q1}
\end{figure}

\begin{figure}
    \centering
\includegraphics[width=0.8\linewidth]{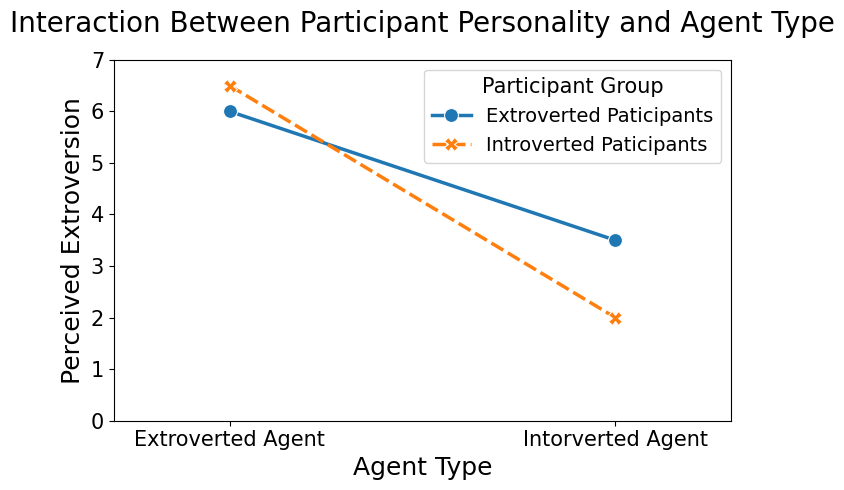}
    \caption{Interaction between Participant Personality and Agent Type on Perceived Extroversion.}
    \label{fig:interaction}
\end{figure}

\subsubsection{Perceived Personality} 

We conducted a 3-way mixed ANOVA to examine the effects of Agent Type (within-subject: Extroverted vs. Introverted), Video Version (within-subject: Version 1 vs. Version 2), and Scenario Type (between-subject: Negotiation vs. Ice-breaking) on the perceived personality scores. 
The analysis revealed a significant main effect of Agent Type ($F = 44.57$, $p < .001$), indicating that extroverted agents were perceived as significantly more extroverted than introverted agents. 
There was no significant main effect of Video Version ($F = 2.99$, $p = .622$), suggesting that perceptions were consistent across the two video versions and not influenced by idiosyncratic video characteristics—supporting the validity of our design. Additionally, there was no significant main effect of Scenario Type, though a marginal trend was observed ($p = .086$).

\begin{table*}[h]
\centering

\resizebox{1\textwidth}{!}{%
\begin{tabular}{l|p{9.0cm}|p{9.0cm}}
\hline
\textbf{Category} & \textbf{Extrovert Agent } & \textbf{Introvert Agent } \\
\hline
\textbf{Cue Influence} & 
Verbal (60\%) > Both Equally (20\%) = Nonverbal (20\%) & 
Verbal (63.3\%) > Both Equally (21.7\%) > Nonverbal (15\%) \\
\hline
\textbf{Nonverbal Cue Influence} & 
Facial Expressions (30.6\%) > Posture (23.5\%) > Gesture = Tone of Voice (22.4\%) > Volume (1.0\%) & 
Facial Expressions = Tone of Voice (32.2\%) > Posture (24.4\%) > Gesture (7.8\%) > Volume (3.3\%) \\
\hline
\textbf{Verbal Cue Influence} & 
Speaking Manner (53.6\%) > Speech Content (37.5\%) > Utterance Length (8.9\%) & 
Speaking Manner (51.8\%) > Speech Content (33.9\%) > Utterance Length (14.3\%) \\
\hline
\end{tabular}
}
\caption{Perceived Influence of Cues on Personality Judgments}
\label{tab:cue-influence}
\end{table*}

\subsubsection{Participant Personality-Perception Interaction} 

We examined whether participants’ own extroversion levels influenced their perception of agent personality, with particular focus on the interaction between Agent Type and Participant Extroversion in a linear regression model.
The analysis revealed significant main effects of Agent Type ($p < .001$) and Participant Extroversion ($p = .009$), as well as a significant interaction between the two factors ($p = .009$).
Figure~\ref{fig:interaction} illustrates this interaction. The blue solid line represents ratings from introverted participants, while the orange dashed line represents ratings from extroverted participants. 
Although both groups rated extroverted agents as more extroverted than introverted agents, the difference between agent types was more pronounced among introverted participants.

\subsubsection{Cue Influence} 

Table~\ref{tab:cue-influence} summarizes participants’ judgments of which cues influenced their perception of agent personality across both agent types. 
Participants rated verbal cues as the most influential for both extroverted and introverted agents (60\% and 63.3\%, respectively)
Among nonverbal cues, facial expressions were most frequently chosen (30.6\% for extroverted, 32.2\% for introverted). 
For extroverted agents, posture (23.5\%), gesture (22.4\%), and tone of voice (22.4\%) followed. 
For introverted agents, tone of voice (32.2\%) was rated equally important as facial expressions, followed by posture (24.4\%). 
Regarding verbal cues, speaking manner was most frequently selected (53.6\% for extroverted, 51.8\% for introverted). 
Chi-square analyses (2\(\times\)2; Agent Type \(\times\) Scenario) were conducted for each cue category (overall cue type, nonverbal cue, verbal cue), and none of the comparisons yielded significant differences. These results suggest that participants’ cue preferences were consistent across both agent type and scenario.

\subsubsection{Behavior Consistency}

We asked participants whether each agent’s verbal and nonverbal behaviors aligned with their perceived personality. 
Most participants perceived the agents’ behaviors as consistent with their intended personalities. For nonverbal behavior, 66.7\% responded “Yes” in Negotiation and 93.3\% in Ice-breaking, with the rest choosing “Somewhat.” For verbal behavior, 73.3\% selected “Yes” in both scenarios, and 26.7\% chose “Somewhat.” Notably, no participants selected “No” for either modality, underscoring the credibility of our personality-aligned behavior generation.
Participants reliably recognized the agents’ intended personalities through their verbal and nonverbal behaviors, supporting the validity of our personality-aligned behavior generation approach.

\section{Discussion}

Our findings demonstrate that both verbal and non-verbal behaviors generated by LLMs reflect the intended personality traits of extroversion and introversion. Through agent-to-agent simulations, we validated that linguistic patterns (as measured by LIWC) and personality predictions from pre-trained models aligned with the assigned personality. Additionally, generated non-verbal behaviors also varied meaningfully based on personality.

The scenario context played a significant role. While personality differences were detectable in both scenarios, they were more pronounced in the Ice-Breaking scenario compared to Negotiation. This difference may be due to the task-oriented nature of Negotiation, where goal-driven behavior could suppress personality expression. 
In contrast, Ice-Breaking involved more casual and socially expressive interactions, allowing clearer differentiation between extroverted and introverted behaviors. This was evident in both LIWC results and non-verbal behavior distributions. Emotionally, anger-related expressions were more common in Negotiation, while expressions associated with happiness and social warmth were more frequent in Ice-Breaking.
Importantly, our framework supports dynamic scenario switching, allowing flexible testing across diverse interaction contexts. This opens possibilities for future work exploring how different situational factors shape personality expression and perception.

Introverted participants rated the introverted agent as more introverted, whereas extroverted participants rated the same agent as less introverted. This pattern may be explained by the ``Assumed Similarity Bias"~\cite{kenny2010similarity}, a tendency for individuals to project their own personality traits onto others. 
As a result, introverted individuals may perceive the introverted agent as more similar to themselves, and likewise for extroverted individuals.
From another perspective, introverted participants appeared to distinguish more clearly between the two agent types. This may suggest that introverted individuals are more sensitive to personality-related cues. While prior work has shown that observer traits systematically influence judgments of others~\cite{human2013targeting}, whether this leads to improved perceptual accuracy remains unclear. Future research should aim to disentangle trait-based accuracy from trait-based bias in personality perception.

\section{Conclusion}

This study investigated whether LLMs can generate both verbal and nonverbal behaviors for embodied virtual agents based on personality traits. 
Focusing on extraversion, we developed a system that combines LLM-generated speech with corresponding nonverbal actions—including facial expressions, body gestures, and voice characteristics—through an action description module. 
Agent-to-agent simulations and user studies confirmed that these behaviors were aligned with the intended personality traits and perceptible to users. 
We also found that verbal behaviors played a greater role than nonverbal behaviors in shaping perceived personality.


Our current user study was based on recorded videos of agent behaviors. As a next step, we plan to conduct real-time interactions between users and virtual agents to further evaluate the system’s effectiveness. 
This would allow us to examine how personality pairing between users and agents influences task-oriented scenarios such as negotiation. 
Such studies could provide deeper insights into the role of personality alignment in shaping interaction outcomes and user experience.
Future work may also investigate the potential of personality pairing in task-driven domains such as negotiation or therapy~\cite{tapus2008user}, where aligning agent behavior with user traits could enhance interaction outcomes.


\bibliographystyle{ACM-Reference-Format}
\bibliography{sample-base}

\end{document}